\documentclass{article}
\usepackage{makeidx}
\usepackage{graphicx}
\usepackage{amsmath}
\usepackage{verbatim}  
\usepackage{tabularx}
\usepackage[comma, authoryear]{natbib} 
\usepackage{url} 

\newcommand{\expectation}[1]{\ensuremath{\overline{P_E}(#1)}}

\newcommand{\estimator}[1]{\ensuremath{\widehat{P_E}(#1) }}
\begin{document}

\title{Surprisingly Rational: \\Probability theory plus noise explains
  biases in judgment.}

\author{\\ \\ Fintan Costello \\ School of Computer Science and
  Informatics\\University College Dublin\\Belfield, Dublin 4,
  Ireland\\ \\ Paul Watts \\Department of Mathematical Physics\\
  National University of Ireland\\Maynooth, Co Kildare, Ireland}

\date{}
\maketitle

\pagebreak
\begin{abstract}
The systematic biases seen in people's probability judgments are
typically taken as evidence that people do not reason about
probability using the rules of probability theory, but instead use
heuristics which sometimes yield reasonable judgments and sometimes
systematic biases.  This view has had a major impact in economics,
law, medicine, and other fields; indeed, the idea that people cannot
reason with probabilities has become a widespread truism.  We present
a simple alternative to this view, where people reason about
probability according to probability theory but are subject to random
variation or noise in the reasoning process.  In this account the
effect of noise is cancelled for some probabilistic expressions:
analysing data from two experiments we find that, for these
expressions, people's probability judgments are strikingly close to
those required by probability theory.  For other expressions this
account produces systematic deviations in probability estimates.
These deviations explain four reliable biases in human probabilistic
reasoning (conservatism, subadditivity, conjunction and disjunction
fallacies).  These results suggest that people's probability judgments
embody the rules of probability theory, and that biases in those
judgments are due to the effects of random noise.
\end{abstract}

\vspace{20mm}
 \textbf{Keywords.} probability; rationality; random variation; heuristics; biases

\pagebreak
\section{Introduction}

The capacity to reason with uncertain knowledge (that is, to reason
with probabilities) is central to our ability to survive and prosper
in ``an ecology that is of essence only partly accessible to
foresight'' \citep{Brunswik1955}.  It is therefore reasonable to
expect that humans, having prospered in such an ecology, would be able
to reason about probabilities extremely well: any ancestors who could
not reason effectively about probabilities would not survive long, and
so the biological basis of their reasoning would be driven from the
gene pool.  Probability theory provides a calculus of chance
describing how to make optimal predictions under uncertainty: taking
the argument one step further, it is reasonable to expect that our
probabilistic reasoning will follow the rules of probability theory.

The conventional view in current psychology is that this expectation
is wrong.  Instead, the dominant position is that
\begin{quote}
In making predictions and judgments under uncertainty, people do not
appear to follow the calculus of chance or the statistical theory of
prediction.  Instead they rely on a limited number of heuristics which
sometimes yield reasonable judgments and sometimes lead to severe and
systematic errors \citep[][p. 237]{TverskyKahneman1973}
\end{quote}
This conclusion is based on a series of systematic and reliable biases
in people's judgements of probability, many identified in the 1970s
and 1980s by Tversky, Kahneman and colleagues.  This heuristics and
biases approach has reached a level of popularity rarely seen in
psychology (with Kahneman recieving a Nobel Prize in part for his work
in this area).  The idea that people do not reason using probability
theory but instead follow various heuristics has been presented both
in review articles describing current psychological research
\citep{gigerenzer2011,ShafirLeBeuf2002}, and in numerous popular
science books summarising this research for the general public
\citep[e.g.][]{ariely2009,kahneman2011}.  This approach has had a
major impact in economics
\citep{camerer2003BehavioralEconomics,behavioralEconomics2003}, law
\citep{korobkin2000law,sunstein2000behavioral}, medicine
\citep{heuristicsBiasesMedicine1987,heuristicsBiasesMedicine2005} and
other fields
\citep{williams2010heuristicsMilitary,hicks2011heuristicsOrthodontics,
  bondt2012StockMarketHeuristics,richards2012heuristicsCIA}.  Indeed
the idea that people cannot reason with probabilities has become a
widespread truism: for example, the Science Gallery in Dublin recently
presented an exhibition on risk which it described as ``enabling
visitors to explore our inability to determine the probability of
everything from a car crash to a coin toss'' ({\it The Irish Times},
Thursday, 11 October 2012).

We have two main aims in this paper: to give evidence against the view
that people reason about probabilities using heuristics, and to give
evidence supporting the view that people reason in accordance with
probability theory, with bias in people's probability estimates being
caused by random variation or noise in the reasoning process.  We
assume a simple model where people estimate the probability of some
event $A$ by estimating the proportion of instances of $A$ in memory,
but are subject to random errors in the recall of instances.  While at first glance it may seem 
that these random errors will result in
``nothing more than error variance centered around a normative
response'' \citep{ShafirLeBeuf2002}, in fact these
random errors cause systematic deviations that push estimates for $P(A)$ 
away from the correct value in a characteristic way.  
In our model these systematic deviations explain various
 biases frequently seen in people's probabilistic reasoning:
conservatism, subadditivity, the conjunction fallacy, and the
disjunction fallacy.  The general patterns of occurrence of these
biases match the predictions of our simple model.

We use this simple model to construct probabilistic expressions that
cancel the bias in estimates for one event against the bias in
estimates for another.  These expressions allow us to test the
predictions of the heuristics view of probabilistic reasoning.  One
such expression involves estimates, for some events $A$ and $B$, of
the individual probabilities $P(A)$ and $P(B)$ and the conjunctive (`and') and disjunctive (`or') probabilities $P(A \wedge B)$  and $P(A \vee B)$.  People's estimates for all four of these
probabilities are typically subject to various forms of bias.  Our
account, however, predicts that when combined in the expression
\begin{eqnarray*}
X_E(A,B)& =& P_E(A)+P_E(B)- P_E(A \wedge B) - P_E(A \vee B)
\end{eqnarray*}
(where $P_E(A)$ represents a person's estimate for $P(A)$, $P_E(B)$
their estimate for $P(B)$, and so on), then the various biases on the
individual expressions will cancel out, and on average $X_E(A,B)$ will
equal $0$ in agreement with probability theory's `addition law' which
requires that
\begin{eqnarray}
\label{additionLaw}
X(A,B)=P(A)+P(B)- P(A \wedge B) - P(A \vee B) &=&0
\end{eqnarray}

Notice that the heuristics view assumes that people estimate
probabilities using heuristics that in some cases yield reasonable
judgments (that is, judgments in accordance with probability theory)
but in other cases lead to systematic biases.  To give evidence
against the heuristics view it is therefore not enough to show that
some of people's probability judgments agree with probability theory
(that is expected in the heuristics view).  Instead, our evidence
against the heuristics view consists of results showing that, even
when people's probability estimates for a set of events are
systematically biased, when those estimates are combined to form
expressions like $X_E$, the results are on average strikingly close to
those required by probability theory.  This cancellation of bias
is difficult to explain in the heuristics view: to explain this
cancellation, the heuristics view would require some way of ensuring
that, when applying heuristics to estimate the probabilities
$P(A)$, $P(B)$, $P(A \wedge B)$ and $P(A \vee B)$ individually,
 the biases produced in those $4$ probabilities are precisely calibrated to give overall
cancellation.  Further, to `know' that the bias in these four
probabilities should cancel out in this way requires access to the
rules of probability theory (as embodied in the addition law in this
case).  Since the heuristics view by definition does not follow the
rules of probability theory, it does not have access to these rules and so
has no reason to produce this cancellation.
 
We also use this model to construct a series of expressions where all
but one `unit' of bias is cancelled; our model predicts that the level
of bias when people's responses are combined in these expressions
should on average have the same constant value.  Our experimental
results confirm this prediction, showing the same level of bias across
a range of such expressions.  Together, these results demonstrate that
when noise in recall is cancelled, people's probability estimates
follow the rules of probability theory and thus suggest that biases in
those estimates are due to noise.  These results are the main
contribution of our work.

Note that our evidence against the view that people use heuristics to
estimate probabilities is not based on the fact that our model
explains the four biases mentioned above (there are many other biases in the
literature which our model does not address; see \citet{Hilbert2012}
for a review).  Instead, the point is that our experimental results
show that the basic idea behind the heuristics view (that people do
not follow the rules of probability theory) is contradicted when we
use our simple model to cancel the effects of noise.
 
\subsection{Bayesian models of reasoning}

We are not alone in arguing that people reason in accordance with
probability theory.  Though ``the bulk of the literature on adult
human reasoning'' goes against this view \citep{cesana2012}, in recent
years various groups of researchers have suggested that people follow
mathematical models of reasoning based on Bayesian inference, a
process for drawing conclusions given observed data in a way that
follows probability theory.  Bayesian inference applies to conditional
probabilities such as the probability of some conclusion $H$ given
some evidence $E$: $P(H|E)$.  In Bayesian models these conditional
probabilities are computed according to Bayes' theorem
\begin{eqnarray*}
P(H|E)&=& \frac{P(H) \times P(E|H)}{P(E)}
\end{eqnarray*}
and so the value of the conditional probability $P(H|E)$ depends on
the value of the `prior' $P(H)$ (the probability of $H$ being true
independent of the evidence $E$) and on the value of the `likelihood
function' $P(E|H)$ (the probability of seeing evidence $E$ given that
the hypothesis $H$ is true).

The status of these Bayesian models is currently controversial.  On
one hand, close fits between human responses and Bayesian models have
been demonstrated in domains as diverse as categorisation, naive
physics, word learning, vision, logical inference, motor control and
conditioning \citep[see e.g.][]{tenenbaum2011,
  chaterTennenbaumYuille2006,oaksford2007}, leading researchers to
conclude that ``everyday cognitive judgements follow [the] optimal
statistical principles'' of probability theory \citep{griffiths2006}.
On the other hand, critics have pointed out a range of problems with
this Bayesian approach
\citep{bowers2012,marcus2013,eberhardt2011,jones2011,Endress2013}.
For example, the estimation of priors and likelihood functions in
Bayesian models is problematic: there are ``too many arbitrary ways
that priors, likelihoods etc.  can be altered in a Bayesian theory
post hoc.  This flexibility allows these models to account for almost
any pattern of results'' \citep{bowers2012}.

It is important to stress that our approach is not connected to this
Bayesian view.  Our model applies only to the estimation of `simple'
probabilities such as the probability of some event $P(A)$, and does
not involve Bayes' theorem or conditional probabilities of any form.
Neither does our model involve parameter estimation, priors, or
likelihood functions.  Equally, our results showing that people's
probability estimates follow the requirements of probability theory
when noise is cancelled do not imply that people follow Bayes' theorem
when estimating conditional probabilities: Bayes' theorem is
significantly more complex than the simple probabilities we consider.

\subsection{Overview} 

In the first section of the paper we present our model and show how it
can explain the observed biases of conservatism, subadditivity, the
conjunction fallacy, and the disjunction fallacy.  In this section we
also discuss other accounts for these biases, some of which are also
based on noise \citep[see][]{Costello2009a,Costello2009b,
  ErevWallstenBudescu1994, Hilbert2012, NilssonWinmanJuslin2009,
  juslin2009, MinervaDM}.  The crucial difference between our account
and others is that our account makes specific and testable predictions
about the degree of bias in probabilistic expressions, and about
expressions where that bias will vanish.  In the second and third
sections we present our model's predictions and describe two
experimental studies testing and confirming these predictions.  In the
final sections we give a general discussion of our work.

\section{Probability estimation with noisy recall}

We assume a rational reasoner with a long-term episodic memory that is
subject to random variation or error in recall, and take $P_E(A)$ to
represent a reasoner's estimate of the probability of event $A$.  We
assume that long-term memory contains $m$ episodes where each recorded
episode $i$ holds a flag that is set to $1$ if $i$ contains event $A$
and set to $0$ otherwise, and the reasoner estimates the probability
of event $A$ by counting these flags.

We assume a minimal form of transient random noise, in which there is
some small probability $d$ that when some flag is read, the value
obtained is not the correct value for that flag.  We assume that this
noise is symmetric, so that the probability of $1$ being read as $0$
is the same as the probability of $0$ being read as $1$.  We also
assume a minimal representation where every type of event, be it a
simple event $A$, a conjunctive event $A \wedge B$, a disjunctive
event $A \vee B$, or any other more complex form, is represented by
such a flag, and where every flag has the same probability $d$ of
being read incorrectly.  (We stress here that this type of sampling
error is only one of many possible sources of noise.  While we use
this simple form of sampling error to motivate and present our model,
our intention is to demonstrate the role of noise -- from whatever
source -- in causing systematic biases in probability estimates.)

We take $C(A)$ to be the number of flags marking $A$ that were read as
$1$ in some particular query of memory, and $T_A$ be the number of
flags whose correct value is actually $1$.  Our reasoner computes an
estimate $P_E(A)$ by querying episodic memory to count all episodes
containing $A$ and dividing by the total number of episodes, giving
\begin{eqnarray}
\label{eq:estimate}
P_E(A) &=& \frac{C(A)}{m}
\end{eqnarray}
Random error in recall (and hence in the value of $C(A)$) means that
$P_E(A)$ varies randomly: sampling $P_E(A)$ repeatedly will produce a
series of different values, varying due to error in recall.  We assume
that this estimation process is the same for every form of event: a
probability estimate for a simple event $A$ is computed from the
number of flags marking $A$ that were read as $1$ in some particular
query of memory, a probability estimate for a conjunctive event $A
\wedge B$ is computed from number of flags marking $A \wedge B$ that
were read as $1$ in some particular query of memory, and so on.

We take $P(A)$ to represent the `true' judgment of the probability of
$A$: the estimate that would be given if the reasoner that was not
subject to random error in recall and produced estimates in a perfect,
error-free manner.  We take $\expectation{A}$ to represent the
expected value or population mean for $P_E(A)$.  This is the value we
would expect to obtain if we averaged an infinite number these
randomly varying estimates $P_E(A)$.  Finally, we take $\estimator{A}$
to represent a sample mean: the average of some finite set of
estimates $P_E(A)$.  This sample mean $\estimator{A}$ will vary
randomly around the population mean $\expectation{A}$, with the degree
of random variation in the sample mean decreasing as the size of the
sample increases.

For any event $A$ the expected value of $P_E(A)$ is  given by
\begin{eqnarray*}
\expectation{A} &=& \frac{T_A (1-d) + (m-T_{A})d}{m}
\end{eqnarray*}
(since on average $1-d$ of the $T_A$ flags whose value is $1$ will be
read as $1$, and $d$ of the $m-T_{A}$ flags whose value is $0$ will be
read as $1$ ).  Since by definition
$$P(A) = \frac{T_A}{m}$$  
we have
\begin{eqnarray}
\label{eq:PEA}
\expectation{A} &=& P(A)+ d- 2 d  P(A) 
\end{eqnarray}
and the expected value of $P_E(A)$ deviates from $P(A)$ in a way that
systematically depends on $P(A)$. 

Individual estimates will vary randomly around this expected value and
so for any specific estimate $P_E(A)$ where $C(A)$ flags were read as
having a value of $1$, we have
\begin{eqnarray}
\label{eq:PEAerror}
P_E(A) &=& \expectation{A}+ e
\end{eqnarray}
 where
\begin{eqnarray*}
e &=& \frac{C(A)-T_A (1-d) - (m-T_A)d }{m}
\end{eqnarray*}
represents positive or negative random deviation from the expected
value across all estimates.  Note that this error term $e$ does not
introduce an additional source of random error in probability
estimates: it simply reflects the difference between the number of
flags that were read incorrectly when computing the specific estimate
$P_E(A)$ and the the number of flags that are read incorrectly on
average, across all estimates.  

Finally, we can also derive an expression 
for the expected variance in these randomly varying estimates $P_E(A)$. 
The expected variance is equal to $d(1-d)/m$ for all events $A$, $A\wedge B$, $A \vee B$
and so on, and is independent of event probability; see the Appendix for details.

\subsection{Conservatism}

In this section we show how our noisy recall model of probabilistic
reasoning explains a reliable pattern of conservatism seen in people's
probability estimates.

Probabilities range in value between $0$ and $1$.  A large body of
literature demonstrates that people tend to keep away from these
extremes in their probability judgments, and so are `conservative' in
their probability assessments. These results show that the closer
$P(A)$ is to $0$, the more likely it is that $P_E(A)$ is greater than
$P(A)$, while the closer $P(A)$ is to $1$, the more likely it is that
$P_E(A)$ is less than $P(A)$.  Differences between true and estimated
probabilities are low when $P(A)$ is close to $0.5$ and increase as
$P(A)$ approaches the boundaries of $0$ or $1$.  This pattern was
originally seen in research on people's revision of their probablity
estimates in the light of further data \citep{Edwards1968}, and was
later found directly in probability estimation tasks.  This pattern is
sometimes referred to as underconfidence in people's probability
estimates \citep[see][for a
  review]{ErevWallstenBudescu1994,Hilbert2012}.

Conservatism will occur as a straightforward consequence of random
variation in our model.  As we saw in Equation \ref{eq:PEA}, the
expected value of $P_E(A)$ deviates from $P(A)$ in a way that
systematically depends on $P(A)$.  If $P(A)=0.5$ this deviation will
be $0$. If $P(A)<0.5$ then since $d$ cannot be negative we have
$\expectation{A} > P(A) $, with the difference increasing as $P(A)$
approaches $0$.  Since estimates $P_E(A)$ are distributed around
$\expectation{A}$ this means that $P_E(A)$ will tend to be greater
than $P(A)$, with the tendency increasing as $P(A)$ approaches $0$.
Similarly if $P(A)>0.5$ then $\expectation{A} < P(A) $ and estimates
$P_E(A)$ will tend to be less than $P(A)$, with the tendency
increasing as $P(A)$ approaches $1$.  This deviation thus matches the
pattern of conservatism seen in people probability judgments.

\subsubsection{Other accounts}
The idea that conservatism can be explained via random noise is not
new to our account, but is also found in
\cite{ErevWallstenBudescu1994}'s account based on random error in
probability estimates, in the Minerva-DM memory-retrieval model of
decision making \citep{MinervaDM}, and in Hilbert's account based on
noise in the information channels used in probabilistic reasoning
\citep{Hilbert2012}.  The underlying idea in these accounts is similar
to ours.  There is, however, a critical difference: our account
predicts no systematic bias for probabilistic expressions with a
certain form (see Section $3$).
 
\subsection{Subadditivity}

Here we show how our noisy recall model explains various patterns of
`subadditivity' seen in people's probability estimates.

Let $A_{1}\ldots A_{n}$ be a set of $n$ mutually exclusive events, and
let $A = A_{1} \vee \ldots \vee A_{n}$ be the disjunction (the `or')
of those $n$ events.  Then probability theory requires that
\begin{eqnarray*}
\sum_{i=1}^nP(A_{i}) &=& P(A) 
\end{eqnarray*}
Experimental results show that people reliably violate this
requirement, and in a characteristic way.  On average the sum of
people's probability estimates for events $A_{1}\ldots A_{n}$ is
reliably greater than their estimate for the probability of $A$, with
the difference (the degree of subadditivity) increases reliably as $n$
increases.  An additional, more specific pattern is also seen: for
pairs of mutually exclusive events $A_{1}$ and $A_{2}$ whose
probabilities sum to $1$ we find that the sums of people's estimates
for $A_{1}$ and $A_{2}$ are normally distributed around $1$, and so on
average this sum is equal to $1$ just as required by probability
theory.  This pattern is sometimes referred to as `binary
complementarity' \citep[see][for a detailed review of these
  results]{TverskyKoehler1994}.

Again, these patterns of subadditivity occur as a straightforward
consequence of random variation in our model.
From Equation \ref{eq:PEA} we have
\begin{eqnarray*}
\sum_{i=1}^n\expectation{A_{i}}&= &\sum_{i=1}^n\left[P(A_{i}) +d-2
d P(A_{i})\right]
\end{eqnarray*}
and using the fact that $P(A_{1})+\ldots + P(A_{n}) = P(A)$ this gives
\begin{eqnarray*}
\label{eq:PEA3}
\sum_{i=1}^n\expectation{A_{i}}&=&
P(A) + n d -2 d P(A)
\end{eqnarray*}
Taking the difference between this expression and that for
$\expectation{A}$ in equation \ref{eq:PEA} we get
\begin{eqnarray*}
\sum_{i=1}^n\expectation{A_{i}}-\expectation{A} &=& (n-1)d
\end{eqnarray*}
and so this difference increases as $n$ increases, producing
subadditivity as seen in people's probability judgments.  In the case
of two mutually exclusive events $A_{1}$ and $A_{2}$ whose
probabilities sum to $1$, from Equation \ref{eq:PEA} we get
\begin{equation*}
\expectation{A_{1}} +\expectation{A_{2}} = P(A_{1}) + d-2 d
P(A_{1}) + P(A_{2}) + d-2 d P(A_{2}) = 1
\end{equation*}
producing binary complementarity as seen in people's judgments.

\subsubsection{Other accounts}
The original account for subadditivity given by
\cite{TverskyKoehler1994} explained the general pattern in terms of an
unpacking process which increased the probability of constituent
events by drawing attention to their components.  This account could
not explain the observed pattern of binary complementarity; to account
for this observation Tversky and Keohler proposed an additional
`binary complementarity' heuristic, which simply stated that there was
no average bias for binary complements.
 
An alternative explanation for subaddivity is given in the Minerva-DM
memory retrieval model of decision
making\citep{MinervaDM,minervaSubadditivity}.  Minerva-DM is a complex
model with a number of different components: it provides a two-step
process for conditional probability judgments, a parameter controlling
the retrieval of items with varying degrees of similarity to the
memory probe (the event whose probability is being judged), a complex
multi-vector representation for stored items in memory, a parameter
controlling the degree of random error in the initial recording of
items in memory, a parameter controlling the degree of random error
causing degradation in stored items, and a parameter controlling the
degree of detail contained in memory probes.  Roughly stated, the
Minerva-DM model estimates the probability of some event by counting
the number of stored items in memory which are similar enough to that
event (whose similarity measure is greater than the similarity
criterion parameter).  Depending on the value of the similarity
criterion, this count will include a number of similar-but-irrelevant
items in addition to items correctly matching the target event.
Because of these similar-but-irrelevant items, the model will give a
probability estimate for the target event that is higher than the true
probability, producing a degree of subaddivity that increases with the
number of component events in the disjunction, just as required.
Note, however, that because this similarity-based account always
increases probability estimates, it cannot explain the observed
pattern of binary complementarity in people's probability judgments,
which can only be explained if one probability is increased and the
other complementary probability is decreased.

More recently, \cite{Hilbert2012} gave an account of subadditivity
based on noise in the information channels used for probability
computation.  Hilbert's model is a very general one, providing for
noise at the initial encoding of data, for noisy degradation of stored
information and for noise during the reading of data from memory.  The
model also specifies three general requirements for the distribution
of noise: that the correct probability is most likely, that noise is
symmetrical around the correct probability and that two binary
complementary probabilities have the same degree of noise. The last of
these requirements allows the model to explain the `binary
complementarity' result observed by \citet{TverskyKoehler1994}.
Beyond these requirements, the model leaves the degree and form of
noise in the system unspecified.  Again, this account is similar to
ours but with the crucial difference that our account predicts no
systematic bias for certain probabilistic expressions.  We give a
further comparison between our model, Hilbert's model and Minerva-DM
in Section $6$.

\subsection{Conjunction and disjunction fallacies}
\label{conjunctionDisjunctionFallacies}
Conservatism and subadditivity both concern averages of people's
probability estimates.  Here we show how our noisy recall model
explains two patterns that involve differences between individual
probability estimates: the conjunction and disjunction fallacies.

Let $A$ and $B$ be any two events ordered so that $P(A) \leq P(B)$.
Then probability theory's `conjunction rule' requires that $P(A \wedge
B)\leq P(A)$; this follows from the fact that $A \wedge B$ can only
occur if $A$ itself occurs.  People reliably violate this requirement
for some events, and commit the `conjunction fallacy' by  
giving probability estimates for conjunctions that
are greater than the estimates they gave for one or other constituent
of that conjunction.    
Perhaps the best-known example of this violation
comes from Tversky \& Kahneman (1983) and concerns Linda:

\begin{quote}
``Linda is 31 years old, single, outspoken, and very bright.  She
  majored in philosophy.  As a student she was deeply concerned with
  issues of discrimination and social justice, and also participated
  in anti-nuclear demonstrations''
\end{quote}

Participants in Tversky \& Kahneman's study read this description and
were asked to rank various statements ``by their probability''.  Two
of these statements were
\begin{quote}
Linda is a bank teller. ($A$)\\
Linda is a bank teller and active in the feminist movement.($A \wedge
B$)
\end{quote}
In Tversky \& Kahneman's initial investigation these two statements
were presented separately, with one group of participants ranking a
set of statements containing $A$ but not $A \wedge B$, and a second
group ranking the same set but with $A$ replaced by $A \wedge B$.  The
results showed that the average ranking given to $A \wedge B$ by the
second group was significantly higher than the average ranking given
to $A$ by the first group, violating the conjunction rule.
 
Note that violation of the conjunction rule can occur in averaged data even when
very few participants are individually committing the conjunction fallacy;
equally, this violation is not necessarily seen in averaged data even
when many participants are individually committing that fallacy.   For this reason,
Tversky \& Kahneman refer to this comparison of averages as an
`indirect' test of the conjunction rule, and describe violations of
that rule in averages as conjunction errors.  Surprised by the results
of their indirect test, Tversky \& Kahneman carried out a series of
increasingly direct tests of the conjunction rule.  In these direct
tests each participant were asked to rank the probability of a set of
statements containing both $A$ and $A \wedge B$.  Tversky \& Kahneman
found that in some cases more than $80\%$ of participants ranked $A
\wedge B$ as more probable than $A$, violating the conjunction rule in
their individual responses.  Tversky \& Kahneman use the term
`conjunction fallacy' to refer only to these direct violations of the
conjunction rule.  Most subsequent studies have focused on similar
direct tests of the conjunction rule in individual probability
estimates (the conjunction fallacy) rather than on indirect tests
comparing averages (the conjunction error).
 
The Linda example is explicitly designed to produce the conjunction
fallacy: this fallacy does not occur for all or even most
conjunctions.  Numerous experimental studies have shown that the
occurrence of this fallacy depends on the probabilities of $A$ and
$B$.  In particular, the greater the difference between $P(A)$ and
$P(B)$, the more frequent the conjunction fallacy is, and the greater
the conditional probability $P(A |B)$, the more frequent the
conjunction fallacy is
\citep{Costello2009a,gavanski1991representativeness,fantino1997conjunction}.

A similar pattern occurs for people's probability estimates for
disjunctions $A \vee B$.  Since $A \vee B$ necessarily occurs if $B$
itself occurs, probability theory requires that $P(A \vee B)\geq P(B)$
must always hold.  People reliably violate this requirement for some
events, giving probability estimates for disjunctions that are less
than the estimates they gave for just one of the constituents.  Just
as for conjunctions, the greater the difference between $P(A)$ and
$P(B)$, and the higher the estimated conditional probability $P(A
|B)$, the higher the rate of occurrence of the disjunction fallacy
\citep{Costello2009b,carlson1989disjunction}.

The observed patterns of conjunction and disjunction fallacy
occurrence arise as a straightforward consequence of random variation
in our model.  The general idea is that our reasoner's probability
estimates $P_E(A)$ and $P_E(A \wedge B)$ will both vary randomly
around their expected values $\expectation{A}$ and $\expectation{A
  \wedge B}$.  This means that, even though $\expectation{A \wedge B}
\leq \expectation{A}$ must hold, there is a chance that the estimate
for $A \wedge B$ will be greater than the estimate for $A$, producing
a conjunction fallacy.  This chance will increase the closer
$\expectation{A \wedge B}$ is to $ \expectation{A}$.

More formally, the reasoner's estimates for probabilites $P_E(A) $ and
$P_E(A \wedge B)$ at any given moment are given by
\begin{eqnarray*}
P_E(A) = \expectation{A} +e_{A}&\mbox{and}&
P_E(A \wedge B) = \expectation{A \wedge B} +e_{A \wedge B}
\end{eqnarray*}
where $e_{A}$ and $e_{A \wedge B}$ represent positive or negative
random deviation from the expected estimate at that time (arising due
to random errors in reading flag values from memory, as in Equation
\ref{eq:PEAerror}).  The conjunction fallacy will occur when $P_E(A) <
P_E(A \wedge B)$, i.e.\ when
\begin{eqnarray*}
\expectation{A} +e_{A} & <& \expectation{A \wedge B} +e_{A \wedge B} 
 \end{eqnarray*}
or, substituting and rearranging, when
\begin{eqnarray}
\label{conjunctionError}
[P(A)- P(A \wedge B)](1-2d) &<&   e_{A \wedge B} - e_{A} 
\end{eqnarray}
holds.  Given that $e_{A \wedge B}$ and $e_{A}$ vary randomly and can
be either positive or negative, this inequality can hold in some
cases.  The inequality is most likely to hold when $P(A)- P(A \wedge
B) $ is low (because in that situation the left side of the inequality
is low).  Since $P(A \wedge B) = P(A|B) P(B)$, we see that $P(A)- P(A
\wedge B) $ is low when $P(A)$ is low and both $P(A|B)$ and  $P(B)$ are
high (or more strictly: when $P(A)$ is close to $0$,  $P(B)$ is close to $1$, and $P(A|B)$ is close to its maximum possible value of $P(A)/P(B)$).  We thus expect the conjunction fallacy to be most frequent when
$P(A)$ is low and $P(A|B)$ and $P(B)$ are both high.  This is just the
pattern seen when the conjunction fallacy occurs in people's
probability estimates.

Reasoning in just the same way for disjunctions, we see that the
disjunction fallacy will occur when
\begin{eqnarray*}
\expectation{B} +e_{B} & >& \expectation{A \vee B} +e_{A \vee B} 
\end{eqnarray*}
or, substituting and rearranging as before, when
\begin{eqnarray*}
[P(A \vee B) -P(B)](1-2d) &<&e_{B}- e_{A \vee B}
\end{eqnarray*}
holds.  But from probability theory, we have the identity
\begin{eqnarray*}
P(A \vee B) -P(B)& =& P(A)- P(A \wedge B)
\end{eqnarray*}
and substituting we see that the disjunction fallacy will occur when
\begin{eqnarray}
\label{disjunctionError}
[P(A)- P(A \wedge B)](1-2d) &<&    e_{B}- e_{A \vee B} 
\end{eqnarray}
and so, just as with the conjunction fallacy, we expect the
disjunction fallacy to be most frequent when $P(A)$ is low and
$P(A|B)$ and $P(B)$ are both high.  Again, this is just the pattern
seen when the disjunction fallacy occurs in people's probability
estimates.

Note that in our model there is an upper limit on the expected rate of
conjunction fallacy occurrence of $50\%$, which occurs when
$\expectation{A \wedge B}=\expectation{A}$: estimates $P_E(A)$ and
$P_E(A \wedge B)$ are distributed around the same population means,
and so the chance of getting $P_E(A) < P_E(A \wedge B)$ is the same as
the chance of getting $P_E(A)>P_E(A \wedge B)$.  The same limit holds
for the disjunction fallacy, and for the same reason.  This limit
occurs because our simple model assumes (somewhat unrealistically)
that the degree of error in recall for examples of $A$ is the same as
the degree of error in recall for for $A \wedge B$ (extensions of our
model which allow for different levels of error in recall for $A$ and $A \wedge B$ 
would not impose this limit).  As we see in the next section,
however, experiments which control for various extraneous factors
typically give conjunction fallacy rates which are consistent with
this limit.

\subsection{The reality of the conjunction fallacy}

A number of researchers have attempted to `explain away' the
conjunction fallacy by pointing to possible flaws in Tversky \&
Kahneman's Linda experiment which may have led participants to give
incorrect responses.  One argument in this line is to propose that the
fallacy arises because participants in the experiments understand the
word `probability' or the word `and' in a way different from that
assumed by the experimenters. A related tactic is to propose that the
fallacy occurs because participants, correctly following the
pragmatics of communication in their experimental task, interpret the
single statement $A$ as meaning $A \wedge \neg B$ ($A$ and not $B$).
Evidence against these proposals comes from experiments using a
betting paradigm, where the word `probability' is not mentioned and
where `and' is demonstrably understood as meaning conjunction, and
experiments where participants are asked to choose among three
different options $A$, $A \wedge B$ and $A \wedge \neg B$.
Conjunction fallacy rates are typically reduced in these experiments
(to between $10\%$ and $50\%$, as compared to the greater than $80\%$
rate seen in Tversky \& Kahneman's Linda experiment), but remain
reliable \citep[see, for
  example,][]{SidesOshersonBoniniViale2002,TentoriBoniniOsherson2004,
wedell2008}.

Another approach is to explain away the conjunction fallacy by arguing
that Tversky \& Kahneman's probabilistic ranking task is not in a form
that is suitable for people's probabilistic reasoning mechanisms,
which (in this argument) are based on representations of frequency.
The suggestion is that if participants were asked to estimate the
frequency with which constituent and conjunctive statements are true,
the conjunction fallacy should vanish \citep{hertwig1999}.  In the
Linda task, this frequency format could involve giving participants a
story about a number of women who fit the description of Linda, then
asking them to estimate `how many of these women are bank tellers' and
`how many of these women are bank tellers and are active in the
feminist movement'.  However, studies have repeatedly shown that while
the occurrence of the conjunction fallacy declines in frequency format
tasks (typically going from a rate greater than $80\%$ in Linda tasks
to a rate between $20\%$ and $40\%$ in frequency format tasks) the
fallacy remains reliable and does not disappear
\citep{mellers2001,fiedler1988,chase1998}.  Indeed, Tversky \&
Kahneman's original 1983 paper examined the role of frequency
formulations in the conjunction fallacy, and found that the fallacy
was reduced but not eliminated by that formulation.\footnote{Note,
  however, that conjunction errors (that is, violations of the
  conjunction rule in indirect tests on averages rather than in direct
  tests on individual responses) can be eliminated by this frequency
  formulation \citep{mellers2001}.}  Taken together, these results
suggest that the conjunction fallacy rates of $80\%$ and above found
by Tversky \& Kahneman are artifically high because of various
confounding factors: studies of the conjunction fallacy that eliminate
these factors give fallacy rates that are generally around $50\%$ or
lower, in line with our model's expectations.
 
\subsubsection{Other accounts}
A large and diverse range of accounts have been proposed for the
conjunction and disjunction fallacies.  Tversky \& Kahneman's original
proposal explained these fallacies in terms of a representativeness
heuristic, in which probability is assessed in terms of the degree to
which an instance is representative of a (single or conjunctive)
category.  Under Tversky \& Kahneman's interpretation, in the Linda
example people gave a higher rating to the conjunctive statement
because the instance Linda was more representative of (that is, more
similar to members of) the conjunctive category `bank-teller and
active-feminist' than the single category `bank-teller'.

Although the representativeness heuristic remains the routine
explanation of the conjunction fallacy in introductory textbooks, a
number of experimental results give convincing evidence against this
account.  Notice that the representativeness heuristic only applies
when a question asks about the probability of membership of an
instance in a conjunctive category, and only applies when knowledge
about representative members of that category is available.  Evidence
against representativeness comes from results showing that the
conjunction fallacy occurs frequently when these requirements do not
hold.  For example, a series of studies by Osherson, Bonini and
colleagues have shown that the conjunction fallacy occurs frequently
when people are asked to bet on the occurrence of unique future
events: such bets are not questions about membership of an instance in
a category, and so representativeness cannot explain the occurrence of
the fallacy in these cases \citep{sides2002,tentori2004,bonini2004}.
\citet{gavanski1991representativeness} found that the conjunction
fallacy occurred frequently when people are asked about categories for
which no representativeness information is available (questions about
imaginary aliens on other planets).
\citet{gavanski1991representativeness} also found that the fallacy
occurred frequently when the probability question was not about the
membership of an instance in a conjunctive category, but about the
membership of two separate instances in two separate single categories
(rather than asking about the probability of Linda being a bank teller
and active feminist, such questions might ask about the probability of
Bob being a bank teller and Linda being an active feminist).  Again,
representativeness cannot explain the occurrence of the conjunction
fallacy for such questions \citep[see][for a review of research in
  this area]{NilssonWinmanJuslin2009}.

The Minerva-DM account gives an alternative explanation for the
conjunction fallacy that is based on the role of similarity in
retrieval \citep{MinervaDM}.  Minerva-DM estimates the probability of
some event by counting the number of stored items in memory whose
similarity to the probe event is greater than the similarity criterion
parameter.  For a conjunction $A \wedge B$, stored items that are
members of $A$ alone or members of $B$ alone can be similar enough to
the conjunction $A \wedge B$ to be (mistakenly) counted as examples of
that conjunction.  If there are a large number of such
similar-but-irrelevant items, the conjunctive probability estimate
$P_E(A\wedge B)$ may be higher than the lower constituent probability
$P_E(A)$, producing a conjunction fallacy response. Note, however,
that because this similarity-based account always increases
probability estimates, it cannot explain the disjunction fallacy
(which occurs when a disjunctive probability estimate is lower than
one of its constituent probabilities).

Other accounts have been proposed where people compute conjunctive
probabilities $P_E(A \wedge B)$ from consituent probabilities $P_E(A
)$ and $P_E(A )$ using some equation other than the standard equation
of probability theory.  In early versions of this approach the
conjunctive probability was taken to be the average of the two
constituent probabilities
\citep{fantino1997conjunction,carlson1989disjunction}.  This averaging
approach does not apply to disjunctive probabilities.  More recently
Nilsson, Juslin and colleagues
\citep{NilssonWinmanJuslin2009,juslin2009} have proposed a more
sophisticated `configural cue' model where conjunctive probabilities
are computed by a weighted average of constituent probability values,
with a greater weight given to the lower constituent probability, and
where disjunctive probabilities are computed by a weighted average
with greater weight given to the higher constituent probability.

Since the average of two numbers is always greater than the minimum of
those two numbers and less than the maximum (except when the numbers
are equal), these averaging accounts predict that the conjunction
fallacy will occur for almost every conjunction (except when the two
constituents have equal probabilities).  This is clearly not the case,
however: there are many conjunctions for which these fallacies occurs
rarely if at all. To address this problem, Nilsson et al.'s model also
includes a noise component which randomly perturbs conjunctive
probability estimates, sometimes moving the conjunctive probability
below the lower constituent probability and so eliminating the
conjunction fallacy for that estimate.  This model thus predicts that
fallacy rates should be inversely related to the degree of random
variation in people's probability judgments, with fallacy rates being
highest when random variation is low and lowest when random variation
is high. This contrasts with our account, which predicts that fallacy
rates should be high when random variation is high and low when random
variation is low. We assess these competing predictions in Section
\ref{randomVariationAnddFallacyRates}.

Finally, we should mention an earlier model for the conjunction
fallacy proposed by one of the authors \citep{Costello2009a}.  Just as
in our current account, this earlier model proposed that people's
probability estimates followed probability theory but were subject to
random variation: this random variation caused conjunction fallacy
responses to occur when constituent and conjunctive probability
estimates were close together.  Apart from that commonality, the two
models are quite different.  Unlike our current account, this earlier
model was not based on the idea of noise causing random errors in
retrieval from memory: instead, that model assumed that estimates
$P_E(A)$ for some event $A$ were normally distributed around the
correct value $P(A)$, and so the average estimate $\expectation{A}$
was equal to the true value $P(A)$.  That earlier model was therefore
unable to account for the patterns of conservatism and subadditivity
seen in people's probability estimates.  Also unlike our current
account, that earlier model assumed that conjunctive and disjunctive
probabilities were computed by applying the equations of probablity
theory to constituent probability estimates, so that $P_E(A \wedge B)=
P_E(A)\times P_E(B |A)$.  This contrasts with the current model, which
computes $P_E(A \wedge B)$ by retrieving episodes of the event $A
\wedge B$ from memory.

\section{Experiment 1}
\label{Expt1}
Our noisy recall model of probability estimation can explain various
patterns of bias in people's probability judgements, and also explain
some specific situations in which those biases vanish (when
probabilities are close to $0.5$, for conservatism; and when two
complementary probabilities sum to $1$, for subadditivity).  We now
present a third situation in which our model predicts that bias will
disappear.

Consider an experiment where we ask people to estimate, for any pair
of events $A$ and $B$, the probabilities of $A$, $B$, $A \wedge B$ and
$A \vee B$.  For each participant's estimates for each pair of events
$A$ and $B$, we can compute a derived sum
\begin{eqnarray*}
X_E(A,B) &=& P_E(A)+P_E(B)- P_E(A \wedge B) - P_E(A \vee B)
\end{eqnarray*}
We can make a specific prediction about the expected value of
$X_E(A,B)$ for all events $A$ and $B$: this value will be
\begin{eqnarray*}
 \overline{X_E}(A,B)&=& \expectation{A} +\expectation{B}
-\expectation{A \wedge B} - \expectation{A \vee B}
\end{eqnarray*}
From Equation \ref{eq:PEA} we get
\begin{eqnarray*}
\overline{X_E}(A,B) &= &[P(A)+ d - 2 d P(A)] + [P(B)+ d - 2 dP(B)]\\
&&-[P(A\wedge B)+ d - 2 d P(A \wedge B)]- [P(A \vee B)+d -2
P(A \vee B)]\\
&=&(1-2 d )[P(A)+ P(B)-   P(A \wedge B)- P(A \vee B)]
\end{eqnarray*}
However, probability theory requires $P(A)+ P(B)- P(A \wedge B)- P(A
\vee B) =0$ for all events $A$ and $B$, and so
$\overline{X_E}(A,B)=0$.  Our prediction, therefore, is that the
average value of $X_E(A,B)$ across all pairs of events $A$ and $B$
will be equal to $0$.  Note that this prediction is invariant: it
holds for all pairs of events $A$ and $B$, irrespective of the degree
of co-occurrence or dependency between those events.

What is the distribution of values of $X_E(A,B)$ around this average
of $0$?  Speaking generally, we would expect this distribution to be
unimodal and roughly symmetric around the mean of $0$ for any pair
$A,B$, since the positive and negative terms in the expression
$X_E(A,B)$ are symmetric: $P(A)+P(B)=P(A\wedge B)+P(A \vee B)$.
 
We examined this expectation in detail via Monte Carlo simulation, by
writing a program that simulates the effects of random noise in recall
on probability estimations for a given set of probabilities.  This
program took as input three probabilities $P(A)$, $P(A \wedge B)$ and
$P(\neg A \wedge B)$ ($not A$ and $B$).  The program constructed a
`memory' containing $100$ items, each item containing flags $A$, $B$,
$A \wedge B$ and $A \vee B$ indicating whether that item was an
example of the given event.  The occurrence of those flags in memory
exactly matched the probabilities of the given event as specified by
the three input probabilities (so the occurrence of $B$, for example,
matched the sum of input probabilities $P(A \wedge B)$ and $P(\neg A
\wedge B)$).  This program also contained a noise parameter $d$ (set
to $0.25$ in our simulations); when reading flag values from memory to
generate some probability estimate $P_E(A)$, the program was designed
to have a random chance $d$ of returning the incorrect value.

We carried out this simulation process for a representative set of
values for the input probabilities $P(A)$, $P(A \wedge B)$ and $P(\neg
A \wedge B)$.  These set consisted of every possible assignment of
values from $\{0,0.1,\ldots,0.9,1.0\}$ to each input probability,
subject to the requirement that both $P(A \wedge B) \leq P(A)$ and
$P(\neg A \wedge B) \leq 1.0 - P(A)$ must both hold.  This requirement
ensures that every set of input probabilities was consistent with the
rules of probability theory.  In total there were $286$ sets of input
probabilities that were consistent with these requirements.  For each
such set of input probabilities the program carried out $10,000$ runs,
on each run generating noisy estimates $P_E(A)$, $P_E(B)$, $P_E(A
\wedge B)$ and $P_E(A \vee B)$ and using those estimates to calculate
a value for the expression $X_E(A,B)$.  These runs give us a picture
of the distribution of values of $X_E(A,B)$.
 
\begin{figure}[t]
  \scalebox{.7}[.7]{
      \includegraphics*[viewport= 80 430 550 750]{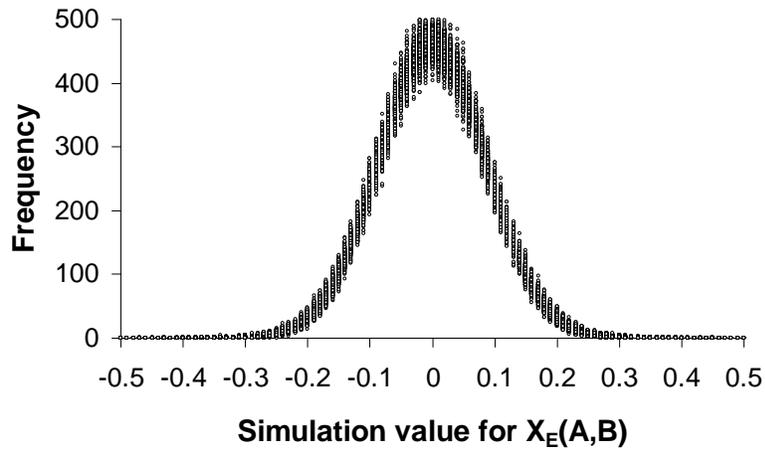}
     } 
      \caption{
  \label{simGraph} Frequency of different values of $X_E(A,B)$ in
        the Monte Carlo simulations.  This scatterplot shows the raw
        frequency of occurrence of different values of $X_E(A,B)$ as
        observed in the Monte Carlo simulations, across runs for a
        range of different probability values $P(A),P(B),P(A \wedge
        B)$ and $P(A \vee B)$ (there were $286$ sets of probability
        values, with $10,000$ $X_E$ values calculated for each set).
        The critical point here is that the distribution of these
        values is essentially the same across different probability
        values: unimodal and symmetrically distributed around $0$.  A
        $-1$ to $+1$ probability scale is used here: note that later
        figures use the $100$ point rating scale from the experiments.
      }
\end{figure}

The distribution of $X_E$ values was essentially the same for all
these sets of input probabilities: unimodal, approximately symmetric,
and centered on $0$, just as expected.  Figure \ref{simGraph} graphs
the frequency distributions of all $X_E$ values across all probability
sets.  Given that this distribution appears to be essentially
independent of the probability values used in our simulations, our
prediction is that in an experiment, the distribution of $X_E(A,B)$
across all pairs of events $A$ and $B$ will be unimodal and
approximately symmetric around the mean of $0$.

One possible concern with this simulation comes from the common
observation that, when estimating probabilities, participants tend to
respond in units that are multiples of $0.05$ or $0.10$
\citep{budescu1988,wallsten1993,ErevWallstenBudescu1994}.  To test the
impact of this rounding, we modified our simulation program to include
a rounding parameter $u$ such that each calculated probability
estimate was rounded to the nearest unit of $u$.  We ran the
simulation as before but with $u$ set to $0.01$, $0.05$ and $0.1$, and
examined the distribution of $X_E$ values for each run: the
distribution was essentially the same as that shown in Figure
\ref{simGraph}, confirming our original simulation results.

\subsection{Testing the predictions}
We tested these predictions using data from an experiment on
conjunction and disjunction fallacies \citep[Experiment $3$
  in][]{Costello2009b}.  The original aim of this experiment was to
examine an attempt by Gigerenzer to explain away the conjunction
fallacy as a consquence of people being asked to judge the probability
of one-off, unique events \citep{gigerenzer1994}.  Gigerenzer argued
that from a frequentist standpoint the rules of probability theory
apply only to repeatable events and not to unique events, and so
people's deviation from the rules of probability theory for unique
events are not, in fact, fallacious.  To assess this argument, the
experiment examined the occurrence of fallacies in probability
judgements for conjunctions and disjunctions of canonical
repeatedly-occurring events: weather events such as `rain', `wind' and
so on.  Contrary to Gigerenzer's argument, participants in these
experiments often committed conjunction and disjunction fallacies;
these fallacies thus cannot be dismissed as an artifact of researchers
using unique events in their studies of conjunctive probability.

This experiment gathered estimates $P_E(A)$, $P_E(B)$, $P_E(A \wedge
B)$ and $P_E(A \vee B)$ from $83$ participants for $12$ pairs $A,B$ of
weather events.  Two sets of weather events (the set `cloudy, windy,
sunny, thundery' and the set `cold, frosty, sleety') were used to form
these pairs.  These sets were selected so that they contained events
of high, medium and low probabilities.  Conjunctive and disjunctive
weather events were formed by pairing each member of the first set
with every member of the second set and placing `and'/`or' between the
elements as required, generating weather events such as `cloudy and
cold', `cloudy and frosty', and so on.  One group of participants
($N=42$) were asked questions in terms of probability, of the form
\begin{itemize}
 \item What is the probability that the weather will be $W$ on a
   randomly-selected day in Ireland?
 \end{itemize}
for some weather event $W$.  This weather event could be a single
event such as `cloudy', a conjunctive event such as `cloudy and cold'
or a disjunctive event such as `cloudy or cold'.  The second group
($N=41$) were asked questions in terms of frequency, of the form
\begin{itemize}
 \item Imagine a set of 100 different days, selected at random.  On
   how many of those 100 days do you think the weather in Ireland
   would be $W$?
 \end{itemize}
where the weather events were as before.  These two question forms
were used because of a range of previous work showing that frequency
questions can reduce fallacies in people's probability judgments; the
aim was to check whether this question form could eliminate fallacy
responses for everyday repeated events.

Participants were given questions containing all single events and all
conjunctive and disjunctive events, with questions presented in random
order on a web browser.  Responses were on an integer scale from $0$
to $100$.  There was little difference in fallacy rates between the
two forms of question, so we collapse the groups together in our
analysis.  There were $996$ distinct conjunction and disjunction
responses in the experiment ($83$ participants $\times 12$
conjunctions): a conjunction fallacy was recorded in $49\%$ of those
responses and a disjunction fallacy in $51\%$.

For every pair of weather events $A,B$ used in the experiment, each
participant gave estimates for the two constituents $A$ and $B$, for
the conjunction $A \wedge B$ and for the disjunction $A \vee B$.  Each
participant gave these estimates for $12$ such pairs.  For each
participant we can thus calculate the value $X_E(A,B)$ for $12$ pairs
$A,B$, and so across all $83$ participants we have $996$ distinct
values of $X_E(A,B)$.  Our prediction is that the average of these
values will equal $0$ and that these values will be approximately
symmetrically distributed around this average.

\subsection{Results}

\begin{figure}[t]
  \scalebox{.7}[.7]{
      \includegraphics*[viewport= 80 430 550 750]{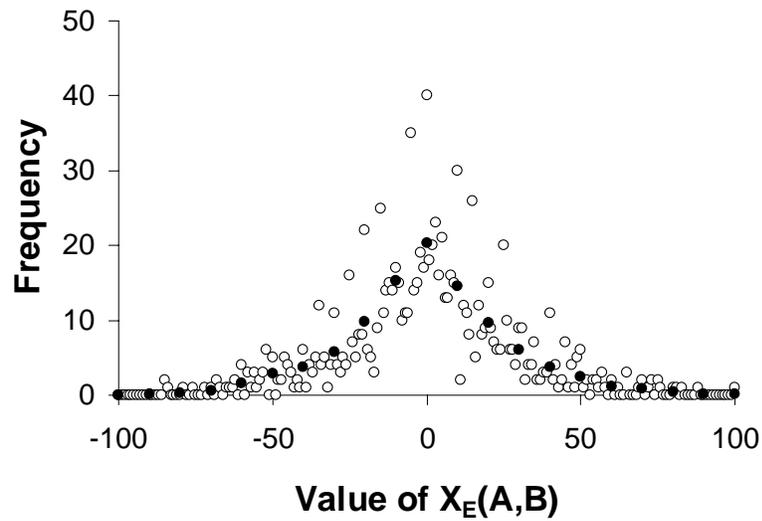}
     } 
      \caption{
  \label{xGraph} Frequency of different values of $X_E(A,B)$ in
        Experiment 1.  This scatterplot shows the raw frequency of
        occurrence of different values of $X_E(A,B)$ as observed in
        the experimental data (hollow circles), and the average
        frequency across grouped values of $X_E(A,B)$ where each group
        contained $10$ values of $X_E(A,B)$ from $v-5 \ldots v+5$ for
        $v$ from $-100$ to $100$ in steps of $10$ (filled circles).
        (The sequence of higher raw frequency values (hollow circles)
        fall on units of $5$, and represent participants rounding to
        the nearest $5$ in their responses). The critical point here
        is that these values are symmetrically distributed around $0$
        as predicted in our model. }
\end{figure}

Figure \ref{xGraph} graphs the raw frequency of occurrence of values
for $X_E(A,B)$ in the experimental data and the average frequency in
groups of those values.  It is clear from the graph that these values
are symmetrically distributed around the mean, just as expected.  The
mean value of $X_E$ was $0.66$ (SD=$27.1$), within $1$ unit of the
predicted mean on the $100$-point scale used in the experiment and
within $0.025$ standard deviations of the predicted mean.  The
predicted mean of $0$ lay within the $99\%$ confidence interval of the
observed mean.  This is in strikingly close agreement with our
predictions.  Note that the sequence of higher raw frequency values
(hollow circles) in Figure \ref{xGraph} fall on units of $5$, and
represent participants' preference for rounding to the nearest $5$
(the nearest unit of $0.05$) in their responses: approximately $55\%$
of all responses were rounded in this way.
   
To examine the relationship between conjunction and disjunction
fallacy rates and $X_E$ values we compared the total number of
conjunction and disjunction fallacies produced by each participant
with the average value of $X_E$ for that participant.  Figure
\ref{versusGraph} graphs this comparison.  There was no significant
correlation between the average value of $X_E$ produced by a
participant and the number of fallacies produced by that participant
($r = -0.1074$, $p = 0.34$).

\subsection{Discussion}
\begin{figure}[t]
  \scalebox{0.7}[0.7]{
      \includegraphics*[viewport= 80 430 550 750]{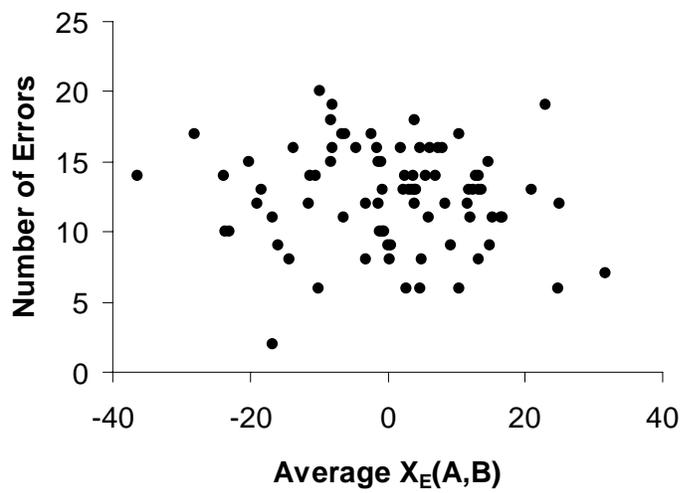}
     } 
      \caption{
  \label{versusGraph} Relationship between conjunction and disjunction fallacies
        and average $X_E(A,B)$ value in Experiment 1.  This
        scatterplot shows the total number of conjunction and
        disjunction fallacies produced by each participant versus the
        average values of $X_E(A,B)$ across all pairs for that
        participant.  The critical point here is that the fallacies
        occur frequently even when $X_E(A,B)=0$.  }
\end{figure}

The above result is based on a specific expression $X_E$ that cancels
out the effect of noise in people's probability judgements.  When
noise is cancelled in this way, we get a mean value for $X_E$ that is
almost exactly equal to that predicted by probability theory.  This
close agreement with probability theory occurs alongside significant
conjunction and disjunction fallacy rates in the same data, with
values of $X_E$ close to zero even for participants with high
conjunction and disjunction fallacy rates (Figure \ref{versusGraph}).
This cancellation of bias is difficult to explain in the heuristics view:
to explain this cancellation, the heuristics view would require some
way of ensuring that, when using heuristics to estimate the $4$
probabilities $P(A)$, $P(B)$, $P(A \wedge B)$ and $P(A \vee B)$ individually, the
various biases in those $4$ estimates are calibrated to give
overall cancellation.  Note that  from both the conservatism results and the binary 
complementarity results described earlier, we know that the bias in 
 estimates $P_E(A)$ and $P_E(B)$ will tend to cancel only when 
$P_E(A) = 1- P_E(B)$ (that is, when $A$ and $B$ are complementary).  
For the heuristics account to explain cancellation of bias 
across the $4$ terms in $X_E$, therefore, it is not enough to say that people 
overestimate $P(A \wedge B)$ and underestimate $P(A \vee B)$: it is 
necessary to calibrate the varying degrees of bias affecting all $4$ probability estimates for $P(A)$, $P(B)$, $P(A \wedge B)$ and $P(A \vee B)$.   

Further, to `know' that the bias in these $4$
probabilities should cancel out in this way requires access to the
rules of probability theory (as embodied in the addition law).  These
results therefore show that that people follow probability theory when
judging probabilities, and that the observed patterns of bias are due
to the systematic distorting influence of noise: when distortions due
to noise are cancelled out as in expression $X_E$, no other systematic
bias remains.

In the next section we describe a new experiment re-testing this
result and testing similar predictions for a range of other
expressions.

\section{Experiment 2}
 
Our prediction for the derived expression $X_E$ holds because the
associated expression $X$ is identically $0$, and because there are an
equal number of positive and negative terms in the expression (these
two requirements are necessary to cancel out the $d$ or noise terms in
the expression).  We now give another expression where these
requirements both hold, and so for which the same prediction follows.

Consider an experiment where we ask people to estimate, for any pair
of events $A$ and $B$, the probabilities of $A$, $B$, $A \wedge B$, $A
\vee B$, $A \wedge \neg B$ ($A$ and not $B$) and $B \wedge \neg A$
($B$ and not $A$).  One derived sum which involves these probabilities
is
\begin{eqnarray*}
Y_E(A,B) &=& P_E(A)+P_E(B \wedge \neg A)-  P_E(B)-P_E(A \wedge \neg B)
\end{eqnarray*}
whose expected value will be
\begin{eqnarray*}
\overline{Y_E}(A,B)&=& \expectation{A} + \expectation{B \wedge \neg A}
-\expectation{B} - \expectation{A \wedge \neg B}
\end{eqnarray*}
From Equation \ref{eq:PEA} we get 
\begin{eqnarray*}
\overline{Y_E}(A,B)&= &[P(A)+ d - 2 d P(A)]+ [P(B \wedge \neg A)+ d
- 2 d P(B \wedge \neg A)]\\&&- [P(B)+ d - 2 d P( B)] - [P(A \wedge \neg
B)+ d - 2 P(A \wedge \neg B)]\\
&=& (1-2 d )[P(A)+ P(B \wedge \neg A)-   P( B)-   P(A \wedge \neg B)]
\end{eqnarray*}
However, probability theory requires $ P(A)+ P(B \wedge \neg A)= P(B)+
P(A \wedge \neg B) $ (because each side of the expression is equal to
$P(A \vee B)$) and so we have $P(A)+ P(B \wedge \neg A) - P(B)- P(A
\wedge \neg B) =0 $ for all events $A$ and $B$, and again we predict
that the average value of $Y_E$ across all participants and event
pairs will equal to $0$.  Since the positive and negative terms in the
expression $Y_E(A,B)$ are symmetric (just as in $X_E$), we again
expect values for $Y_E(A,B)$ to be symmetrically distributed around
this mean, just as with $X_E$ (this prediction is supported by Monte
Carlo simulations similar to those described earlier).  Finally, since
both $X_E$ and $Y_E$ have the same mean of $0$, we predict that the
larger combined set of all values of $X_E$ and $Y_E$ across all
participants and event pairs will also have an mean of $0$, and will
be symmetrically distributed around that mean.

We can also consider other derived sums whose values in probability
theory are $0$, but where there is not an equal number of positive and
negative terms in the expression (and so not all $d$ or noise terms
will be cancelled out).  Four such expressions are
\begin{eqnarray*}
Z1_E(A,B) &=& P_E(A)+P_E(B \wedge \neg A)-P_E(A \vee B)\\
Z2_E(A,B) &=& P_E(B)+P_E(A \wedge \neg B)-P_E(A \vee B)\\
Z3_E(A,B) &=& P_E(A \wedge \neg B)+P_E(A \wedge B) -P_E(A)\\
Z4_E(A,B) &= &P_E(B \wedge \neg A)+P_E(A \wedge B) -P_E(B)
\end{eqnarray*}
For the first expression $Z1_E(A,B)$, Equation \ref{eq:PEA} gives
\begin{eqnarray*}
\overline{Z1_E}(A,B)& = & [P(A)+ d - 2 d P(A)] + [P(B \wedge \neg A)+
  d - 2 d P(B \wedge \neg A)] \\&& - [P(A \vee B)+ d - 2 P(A \vee B)]
\\ &= & (1-2d) [P(A)+ P(B \wedge \neg A)- P(A \vee B)]+d \\ &=& d
\end{eqnarray*}
since from probability theory $P(A)+ P(B \wedge \neg A)- P(A \vee
B)=0$ for all $A$ and $B$.  We get the same result for the expressions
$Z2$, $Z3$ and $Z4$, and so we have expected values of
\begin{eqnarray*}
\overline{Z1_E}(A,B) = \overline{Z2_E}(A,B) = \overline{Z3_E}(A,B) =
\overline{Z4_E}(A,B) &=&d
\end{eqnarray*}
 for all pairs $A,B$.  Our prediction, therefore, is that expressions
 $Z1 \ldots Z4$ should all have the same average value in our
 experiment.
 
Two other such derived sums are
\begin{eqnarray*}
Z5_E(A,B) &=& P_E(A \wedge \neg B) + P_E(B \wedge \neg A)+P_E(A \wedge
B) -P_E(A \vee B)\\
Z6_E(A,B) &=& P_E(A \wedge \neg B) + P_E(B \wedge \neg A)+P_E(A \wedge
B)+P_E(A \wedge B)\\
&&-P_E(A)-P_E(B)
\end{eqnarray*}
Similiar computations for these expressions tell us that $Z5$ will
have an expected value of $ \overline{Z5_E}(A,B)=2d$ and $Z6_E$ will
have an expected value $ \overline{Z6_E}(A,B)=2d$ for all pairs $A,B$.
Our prediction, therefore, is that expressions $Z5$ and $Z6$ should
have the similar average values in our experiment, and that this
average should be twice the average for $Z1 \ldots Z4$.  Note that
since these expressions are not symmetric (all have `leftover' $d$
terms) we do not expect the values of these expressions to be
symmetrically distributed.

These last predictions are somewhat similar to the subadditivity
results described earlier, in that both involve leftover $d$ terms.
The subadditivity results only applied to disjunctions of exclusive
events (events that did not co-occur).  The current predictions are
more general in that they hold for all pairs of events $A$ and $B$,
irrespective of the degree of co-occurrence or dependency between
those events.

\subsection{Method}

\textit{Participants.} Participants were $68$ undergraduate students
at the School of Computer Science and Informatics, UCD, who
volunteered for partial credit.
 
\textit{Stimuli.} This experiment gathered people's estimates for
$P(A)$, $P(B)$, $P(A \wedge B)$, $P(A \vee B)$, $P(A \wedge \neg B)$
and $P(B \wedge \neg A)$ for $9$ different pairs $A,B$ of weather
events such as `rainy',`windy' and so on.  We constructed sets of
three weather events each (the set `cold, rainy, icy' and the set
`windy, cloudy, sunny'), selected so that each set contained events of
high, medium and low probabilities.  Note that some of these pairs
have positive dependencies (it is more likely to be rainy if it is
cloudy), some had negative dependencies (it is less likely to be cold
if it is sunny), and others were essentially independent: our
predictions apply equally across all cases.

Conjunctive and disjunctive weather events were formed from these sets
by pairing each member of the first set with every member of the
second set and placing `and'/`or' between the elements as required,
generating weather events such as `cold and windy', `cold or cloudy'
and so on. Weather events describing conjunctions with negation were
constructed by pairing each member of the first set with every member
of the second set, taking the two possible orderings of the selected
elements, and for each placing `and not' between the elements in each
ordering.  This generated events such as `cold and not windy' and
`windy and not cold' for each pair of events.

\textit{Procedure.}  Participants judged the probability of all single
events and all conjunctions, disjunctions and conjunctions with
negations.  Questions were presented in random order on a web browser.
One group of participants ($N=35$) were asked questions in terms of
probability, of the form
\begin{itemize}
 \item What is the probability that the weather will be $W$ on a
   randomly-selected day in Ireland?
 \end{itemize}
for some weather event $W$.  This weather event could be a single
event such as `cloudy', a conjunctive event such as `cloudy and cold',
a disjunctive event such as `cloudy or cold', or a conjunction and
negation event such as `cloudy and not cold' or `cold and not cloudy'.
The second group ($N=33$) were asked questions in terms of frequency,
of the form
\begin{itemize}
 \item Imagine a set of 100 different days, selected at random.  On
   how many of those 100 days do you think the weather in Ireland
   would be $W$?
 \end{itemize}
 where the weather events were as before.  Responses were on an
 integer scale from $0$ to $100$.  The experiment took around $40$
 minutes to complete.
 
\subsection{Results}

Two participants were excluded (one because they gave responses of
$100$ to all but $4$ questions and the other because they gave
responses of $0$ to all but $2$ questions), leaving $66$ participants
in total.  There were thus $594$ distinct conjunction and disjunction
responses for analysis in the experiment ($66$ participants $\times 9$
conjunctions): a conjunction fallacy was recorded in $46\%$ of those
responses and a disjunction fallacy in $40\%$.

For every pair of weather events $A,B$ used in the experiment, each
participant gave probability estimates for the two constituents $A$
and $B$, for $A \wedge B$ and $A \vee B$, and for $A \wedge \neg B$
and $B \wedge \neg A$.  Each participant gave these estimates for nine
such pairs.  For each participant we calculated the value $X_E(A,B)$,
$Y_E(A,B)$ and $Z1_E(A,B) \ldots Z6_E(A,B)$ for nine pairs $A,B$, and
so across all $66$ participants we have $594$ distinct values for each
of those expressions.

\subsubsection{Expressions $X_E$ and $Y_E$}
\begin{figure}[t]
\scalebox{.7}[.7]{
\includegraphics*[viewport= 20 400 550 750]{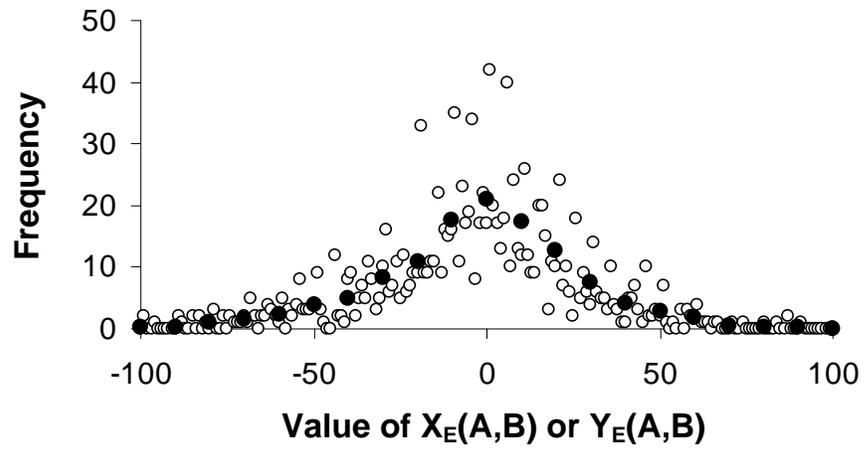}
} 
\caption{
  \label{xGraphExpt2} Frequency of different values of $X_E(A,B)$
and $Y_E(A,B)$ in Experiment 2.  This scatterplot shows the raw
frequency of occurrence of different values of $X_E(A,B)$ and
$Y_E(A,B)$ as observed in the experimental data (hollow circles), and
the average frequency across groups of $10$ values as in Figure $1$.
The critical point here is that these values are symmetrically
distributed around $0$ as predicted in our model. }
\end{figure}

The mean value of $X_E$ was $-3.90$ (SD=$27.7$) and the mean value of
$Y_E$ was $3.82$ (SD=$30.08$).  Figure \ref{xGraphExpt2} graphs the
raw frequency of occurrence of values of $X_E$ and $Y_E$ in the
experimental data and the average frequency in groups of those values,
as in Figure \ref{xGraph}.  It is clear that these values are again
unimodal and symmetrically distributed around their mean, as
predicted.  Averaging across all values of $X_E$ and $Y_E$ we get a
mean of $-0.01$ (SD=$29.2$); the predicted mean of $0$ lies with the
$99.9\%$ confidence interval of this observed mean.  (Again, the
sequence of higher raw frequency values (hollow circles) in Figure
\ref{xGraphExpt2} fall on units of $5$, and represent participants'
preference for rounding to the nearest $5$ in their responses:
approximately $55\%$ of all responses were rounded in this way.)
  
To examine the relationship between conjunction and disjunction
fallacy rates and $X_E$ and $Y_E$ values we compared the total number
of conjunction and disjunction fallacies produced by each participant
with the average $X_E$ and $Y_E$ values for that participant.  As in
the previous experiment, there was no significant correlation between
the average values produced by a participant and the number of
fallacies produced by that participant ($r = -0.073$ and $r = -0.018$
respectively); the results showed values of $X_E$ and $Y_E$ close to
zero even for participants with high conjunction and disjunction
fallacy rates.

As before, these cancellation of bias results are difficult for the
heuristics view to explain: they would require some way of ensuring
that, when using heuristics to estimate the $4$ constituent
probablities in $X_E$, and the $4$ constituent probabilities in $Y_E$, 
the resulting biases are precisely calibrated to give overall cancellation.  
Further, to know that the
bias in these probabilities should cancel requires access to the rules
of probability theory, which the heuristics view does not have.

\subsubsection{Expressions $Z1_E,\ldots, Z4_E$}

Recall that in our model estimates for expressions $Z1,\ldots, Z4$
should on average all have the same biased value, equal to $d$ (the
noise rate). Table $1$ gives the average values for these expressions
calculated from participant's probability estimates; it is clear that
these values are closely clustered (all are less than one tenth of an
$SD$ from the mean) just as predicted.

\subsubsection{Expressions $Z5_E$ and $Z6_E$}

Recall that in our model estimates for expressions $Z5_E$ and $Z6_E$
should on average have the same biased value, equal to $2d$ (twice the
noise rate).  Our prediction therefore is that values for $Z5_E$ and
$Z6_E$ should fall close together, and should fall close to twice the
overall mean obtained for expressions $Z1_E, \ldots, Z4_E$ (as in
Table $1$).  Table $2$ gives the average values for these expressions
calculated from participant's probability estimates, and compares with
twice the overall mean of $Z1_E, \ldots, Z4_E$.  It is clear that
these values are closely clustered around that predicted value (both
are less than one-twentieth of an SD from the predicted value, and
their mean is less than $0.001$ SD from that predicted value).

According to our model, the mean values of expressions $Z1_E, \ldots,
Z4_E$ are equal to the average value of $d$, the rate of random error
in recall from memory, and the mean values of expressions $Z5_E$ and
$Z6_E$ are equal to twice that value.  This raises the interesting
possiblity of using the values of expressions $Z1_E, \ldots, Z6_E$ for
a given participant to estimate a value of $d$ for that participant,
and so estimate the degree of variability due to noise in that
participant's probability estimates.  We discuss this possibility in
the next section.

 \subsection{Random variation and fallacy rates across participants}
 \label{randomVariationAnddFallacyRates}
 
In our model the rate of occurrence of the conjunction fallacy is
related to the degree of random variation: if there were no random
variation in participant's estimates the fallacy would never occur,
while if there is a high degree of random variation the fallacy would
occur frequently.  The same prediction applies to the disjunction
fallacy.

In this section we test these predictions using the data from our
Experiment $2$.  In this analyis we use each participant's average
values for expressions $Z1_E \ldots Z6_E$ to estimate a value of $d$,
the rate of random variation in recall for that participant.  For each
participant we can compute $6$ estimates for that participant's value
of $d$, by taking that participant's average value for each expression
$Z1_E, \ldots, Z6_E$ (and dividing the averages for $Z5_E$ and $Z6_E$
by $2$).  To examine the consistency of these estimates, we computed
the pairwise correlation across participants between values of $d$
estimated from each expression.  Every pairwise correlation was
significant at the $p<0.0001$ level, and the average level of
correlation was relatively high (mean $r = 0.79$), indicating that the
values for $d$ estimated for each participant from each of these
expressions were consistent with each other.

Given this consistency we can produce an average estimate for $d$ for
each participant $i$:
\begin{eqnarray*}
d_{i} &= &\frac{Z1_{i}+Z2_{i}+Z3_{i}+Z4_{i}+Z5_{i}/2+Z6_{i}/2}{6}
\end{eqnarray*}
where $Z1_{i}$ represents the average value of the derived sum
$Z1_E(A,B)$ computed from participant $i$'s probability estimates for
the nine pairs $A,B$.  This gives a reasonable measure of the degree
of random variation in recall for that participant.
 
To test our prediction that conjunctive and disjunctive fallacy rates
will rise with the degree of random variation, we measure the
correlation between conjunction and disjunction fallacy rates and the
$d_{i}$ random variation measure, across participants.  There was a
significant positive correlation between conjunction fallacy rates and
the random variation measure ($r=0.57, p<0.0001$) and between
disjunction fallacy rates and the random variation measure ($r=0.43, p
<0.0005$), demonstrating that fallacy rates rise with random variation
as in our model.  This result goes against Nilsson et al.'s model,
which predicts that conjunction and disjunction fallacy rates will
fall with random variation \citep{NilssonWinmanJuslin2009}.

\section{Conjunction error rates in averaged estimates}
\label{averagedEstimates}

In this section we consider the occurrence of the conjunction (and
disjunction) fallacy in values that are produced by averaging across a
set of probability estimates.  Recall that Tversky \& Kahneman's
initial investigation found that the average probability rankings
given to a conjunction $A \wedge B$ by one group of participants was
reliably higher than the average rankings given to a constituent $A$
by another group of participants, producing a conjunction error
\citep{TverskyKahneman1983}.  More recently
\citet{NilssonWinmanJuslin2009} carried out a detailed study on the
occurrence of conjunction and disjunction fallacies in averaged
probability estimates.  In Nilsson et al.'s experiments, participants
assessed the probability of conjunctions, disjunctions, and
constituent events in a `test-retest' format, with each participant
being asked to assess each probability twice, once in block $1$ and
once in block $2$.  Nilsson et al.  calculated the average probability
estimate given by each participant for each constituent, conjunction
and disjunction, and found that conjunction and disjunction fallacies
in the averaged estimates were more frequent than conjunction and
disjunction fallacies in the individual probablity estimates.  In this
section we show that our model can account for this pattern of
results.  We first discuss the factors in our model that cause this
pattern, and then give simulation results demonstrating their
occurrence in the model.
 
Consider a series of repeated experiments where in each experiment we
gather $N$ estimates for $A$ and $A \wedge B$.  For each experiment,
the sample means $\estimator{A}$ and $\estimator{A \wedge B}$
represent averages of the $N$ probability estimates obtained in that
experiment.  Across experiments, these sample means will vary randomly
around their population means $\expectation{A \wedge B}$ and
$\expectation{A}$, with different experiments giving different sample
means, just as individual estimates $P_E(A)$ and $P_E(A \wedge B)$
vary randomly around those same population means.  The conjunction and
disjunction fallacy results described for individual estimates in
Section \ref{conjunctionDisjunctionFallacies}, which depended on this
random variation in individual estimates, thus also apply to sample
means; the only difference between the two situations is that the
degree of random variation in sample means will decline as the sample
size $N$ increases.

The chance of a conjunction fallacy in sample means (that is, the
chance of getting $\estimator{A} < \estimator{A \wedge B}$ in an
experiment) depends on various factors.  One factor is is the number
of individual estimates $N$ being averaged; another is the difference
between the probabilities $P(A)$ and $P(A \wedge B)$ being estimated.
If $P(A \wedge B)$ and $P(A)$ are far apart, then population means
$\expectation{A \wedge B}$ and $\expectation{A}$ will also be far
apart and a conjunction fallacy can only occur when there is a large
degree of variation in the sample means (that is, when $N$ is low).
On the other hand, if $P(A \wedge B)$ and $P(A)$ are close, then the
population means will be close, and a conjunction fallacy can occur
even when the degree of variation in sample means is low (that is,
when $N$ is high).  In other words, as sample size $N$ increases,
conjunction fallacy rates in sample means will decrease, with the rate
of decrease depending on the difference between $P(A)$ and $P(A \wedge
B)$.
  
When $P(A \wedge B) = P(A)$ the situation is different.  Recall that
in our model the distribution of individual probability estimates for
some event depends only on the number of occurrences of that event in
memory (see Equation \ref{eq:estimate}).  When $P(A \wedge B) = P(A)$
the number of occurrences of $A \wedge B$ is the same as the number of
occurrences of $A$, and so the distribution of estimates for $A$ is
identical to the distribution for $A \wedge B$.  This means that
$\estimator{A}$ and $\estimator{A \wedge B}$ have the same random
distribution around the same population mean.  Because population
means and distributions are the same, the chance of getting $
\estimator{A} < \estimator{A \wedge B}$ is exactly the same as the
chance of getting $\estimator{A }> \estimator{A \wedge B}$.  Since the
first of these two possibilities produces a conjunction fallacy in the
averaged data, the chance of getting a conjunction fallacy is
\begin{equation}
\label{eq:Eq}
\frac{1-Eq}{2}
\end{equation}
where $Eq$ represents the chance of getting exactly the same values
for $\estimator{A}$ and $\estimator{A \wedge B}$.  If we assume that
$\estimator{A}$ and $\estimator{A \wedge B}$ are continuous rather
than discrete variables, then $Eq$ is negligible, and we see that when
$P(A \wedge B) = P(A)$ the chance of getting a conjunction fallacy in
averaged data is $0.5$ for all sample sizes $N$.

Consideration of the chance of getting exactly the same values for two
sample means $\estimator{A}$ and $\estimator{A \wedge B}$ brings us to
a third factor influencing conjunction error rates: rounding in
participant responses.  Recall our earlier observation that when
estimating probabilities, participants tend to respond in units that
are multiples of $0.05$ or $0.10$.  For small $N$ this rounding of
estimates produces sample means that are {\em not} continuous
variables, but instead only take on a limited range of values; for
example, if individual estimates are rounded to units of $0.10$, then
for $N=1$ sample means can only have values that are multiples of
$0.10$, for $N=2$ sample means can only have values that are multiples
of $0.05$, for $N=3$ sample means can only have values that are
multiples of $0.03333$, and so on.  This limitation on the range of
possible values for sample means increases the chance of getting
exactly the same values for $\estimator{A}$ and $\estimator{A \wedge
  B}$; that is, increases the value of $Eq$.  $Eq$ is highest when $N$
is small (when there is only a small range of possible values for the
sample means) and declines as $N$ increases.  Since a high value for
$Eq$ means a low conjunction fallacy rate in sample means (Equation
\ref{eq:Eq}), this rounding effect causes the rate of conjunction
fallacies in sample means to increase with increasing sample size $N$.
This rounding effect can thus explain the increase in conjunction
fallacy rate when averaging across multiple estimates that was
observed by Nilsson et al.  The same reasoning applies to
disjunctions.
 
\subsection{Simulation of  Nilsson et al.'s Experiments}
  
To test this explanation for Nilsson et al.'s results we use an
extension of the simulation program described earlier (see Section
\ref{Expt1}).  This extension simulates Nilsson et al.'s Experiment 2,
which directly compared averaged conjunction error rate against
individual conjunction fallacy rate.

The stimuli in Nilsson et al.'s Experiment $2$ consisted of $180$
components, $90$ conjunctions and $90$ disjunctions that were
constructed by randomly pairing those components.  Components were
constructed using a list of $188$ countries: each component consisted
of a proposition stating that a given country had a population greater
than $6,230,780$ (the median population for the list).  For example, a
component could read ``Sweden has a population larger than
6,230,780'': participants in the experiment were asked to indicate
whether they thought that statement was true or false, and to give
their confidence in that judgment on a $5$ point scale going from
$50\%$ to $100\%$.

A unique sample of components was created for each participant by
randomly sampling, with replacement, from the set of components.
Conjunctions and disjunctions were constructed by randomly pairing
components (excluding duplication) so one conjunction could read
``Sweden has a population larger than 6,230,780 \textit{ and } Spain
has a population larger than 6,230,780'': participants were asked to
indicate whether they thought that statement was true or false, and to
give their confidence in that judgment on a $5$ point scale going from
$50\%$ to $100\%$.

Participants' responses were transformed to a $0\%$ to $100\%$ scale
by subtracting from $100$ the confidence rating for those items where
the participants gave a `false' response.  This experiment thus
necessarily rounds participants' responses to units of $10\%$.  The
experiment had a `test-retest' design, where each participant was
asked to estimate the probability for every component, conjunction and
disjunction twice, in two separate blocks.  Nilsson et al.'s primary
result was that there was a higher rate of conjunction and disjunction
fallacy occurrence when estimates were averaged across the two blocks
than there was in the individual blocks alone.

Participant responses in Nilsson et al.'s experiment represent
judgments in the confidence that a given country's population is above
the median (or, for conjunctions, that a pair of countries are above
the median).  To simulate this experiment we start with a
representation of the `true' confidence that a given population is
above the median.  This true confidence is then input to our
simulation program, which models the effect of random error in causing
variation in that confidence.  To mirror Nilsson et al.'s experiment
as closely as possible, we simulate these true confidence
values using a list of $188$ highest country populations from
Wikipedia\footnote{
  \url{http://en.wikipedia.org/wiki/List_of_countries_by_population},
  accessed Feb 20, 2014}, and the median population for those
countries.  To construct simulated confidence judgments analogous to
those given by Nilsson et al.'s participants, we took $p_i$ to
represent the population of country $i$ and $p_m$ to represent the
median population, and reasoned that the greater the difference
between $p_i$ and $p_m$, the greater the confidence there should be in
judging that country $i$ has a population greater (or less) than the
median.  For countries with populations greater than the median we
therefore took
\begin{equation*}
0.5 + \frac{p_i - p_m}{2 \times \max(p_i,p_m)}
\end{equation*}
to represent a simulated measure of confidence in the country's
population being greater than the median.  Similarly, for countries
with populations less than the median we took
\begin{equation*}
0.5 + \frac{p_m - p_i}{2 \times \max(p_i,p_m)}
\end{equation*}
to represent a simulated measure of confidence in the country's
population being less than the median.  Note that both these
confidence measures run from $0.5$ to $1$, just as in Nilsson et al.'s
experiment.  Finally, we transformed these simulated confidence
measures onto a $0$ to $1$ scale by following Nilsson et al.'s
procedure and subtracting from $1$ the confidence measure for
countries with population less than the median.  For every country
$i$, this procedure gave a component probability
\begin{eqnarray*}
P(i>median)& =& 0.5 + \frac{p_i - p_m}{2 \times \max(p_i,p_m)}
\end{eqnarray*}
that corresponds to a simulated measure of confidence that the
population of country $i$ is greater than the median (and where values
less than $0.5$ represent countries whose population is less than the
median).  To construct conjunctive and disjunctive probabilities we
simply applied the probability theory equations for conjunction and
disjunction to those component probabilities, under the assumption
that component probabilities were independent.

On each run our simulation program took as input $180$ randomly
selected `true' confidence judgments for components (values $P(A)$,
$P(B)$) constructed by applying the calculations described above to
$180$ randomly-selected countries, and $90$ conjunctive and
disjunctive confidence judgments ($P(A \wedge B)$ and $P( A \vee B)$)
calculated by applying the equations of probability theory to those
components.  For each set of components, conjunctive and disjunctive
values, the program constructed a `memory' containing $100$ items,
each item containing flags $A$, $B$, $A \wedge B$ and $A \vee B$
indicating whether that item was an example of the given event.  The
rate of occurrence of those flags in memory matched the values
specified by the input probabilities.  The program contained a noise
parameter $d$ (set to $0.25$ as before) and a rounding parameter set
to $u=0.1$ to match the rounding to units of $10\%$ in Nilsson et
al.'s experiment.  To match the test-retest format in Nilsson et al.'s
experiment, the program obtained two separate estimates of $P_E(A)$,
$P_E(B)$, $P_E(A \wedge B)$ and $P_E(A \vee B)$ for each set of input
probabilities.  For each run the program returned the proportion of
conjunction (and disjunction) fallacy responses in the individual
blocks, and the proportion of conjunction (and disjunction) fallacy
responses in the averages across those two blocks.

Each run of the program thus corresponded to a simulated participant
in Nilsson et al.'s experiment.  We ran the program $1000$ times and
compared the proportion of conjunction and disjunction fallacy
responses in the individual blocks against the proportion of
conjunction and disjunction fallacy responses in the averages.  Just
as in Nilsson et al.'s experiment, there was a higher rate of
conjunction fallacy responses in averages ($M=0.19$, SD=$0.04$) than
in individual blocks ($M=0.16$, SD=$0.03$, $t(1998)=18.76$, $p <
0.0001$), and a higher rate of disjunction fallacy responses in
averages ($M=0.19$, SD=$0.04$) than in indidual blocks ($M=0.15$,
SD=$0.03$, $t(1998)=20.13$, $p < 0.0001$). These simulation results
show that our model is consistent with the pattern seen in Nilsson et
al.'s experiment.
  
\section{General Discussion}

We can summarise the main results of our experiments as follows: when
distortions due to noise are cancelled out in people's probability
judgments (as in $X_E$), those judgements are,
on average, just as required by probability theory with no systematic
bias.  This close agreement with probability theory occurs alongside
significant conjunction and disjunction fallacy rates in people's
responses.  This cancellation of bias cannot be explained in the
heuristics view: to explain this cancellation, the heuristics view
would require some way of ensuring that, when applying heuristics to
estimate the various probabilities in expressions like $X_E$, the
biases produced by heuristics are precisely calibrated to give overall
cancellation. 

Note that cancellation in the expression $X_E$ is 
required because of probability theory's addition law (Equation \ref{additionLaw}), which
is equivalent to 
$$P(A \vee B)= P(A)+P(B)-P(A \wedge B)$$
(the probability theory equation for disjunction).  If the 
heuristics view were able to ensure cancellation for 
$X_E$, then that would mean the heuristics were 
 embodying this addition law; or in other words, that 
the heuristics were implementing the probability theory 
equation for disjunction.  However, this undermines the 
fundamental idea of the heuristics view, which is that people 
do not reason according to the rules of probability theory. 
Since the assumption in the heuristics view is that people do 
not follow probability theory when estimating probabilities, 
there is no way, in the heuristics view,  to know 
that the terms in  $X_E$ should cancel.  To put 
this point another way: if  heuristics were selected 
in  some way to ensure cancellation of bias for $X_E$, 
they would no longer be heuristics: they would simply be 
instantiations of probability theory. 
 
Furthermore, our results show that when one noise term is left after
cancellation (as in expressions $Z1_E \ldots Z4_E$), a constant `unit'
of bias is observed in people's probablity judgments, and when two
noise terms are left after cancellation (as in expressions $Z5_E$ and
$Z6_E$), twice that `unit' of bias is observed in people's judgments,
just as predicted in our simple model.  Again, these results cannot be
explained by an account where people estimate probabilities using
heuristics: such an account would not predict agreement in the degree
of bias across such a range of different expressions.  Together, these
results demonstrate that people follow probability theory when judging
probabilities in our experiments and that the observed conjunction and
disjunction fallacy responses are due to the systematic distorting
influence of noise and are not systematically influenced by any other
factor.

It is worth noting that, while our results demonstrate that people's
probability estimates given in our experiments followed probability
theory (when bias due to noise is cancelled out), we do not think
people are consciously aware of the equations of probability theory
when estimating probabilities.  That is evidently not the case, given
the high rates of conjunction and disjunction fallacies in people's
judgments.  Indeed we doubt whether any of the participants in our
experiment were aware of the probablity theory's requirement that our
various derived sums should equal $0$ or would be able to apply that
requirement to their estimations.  Instead we propose that people's
probability judgments are derived from a `black box' module of
cognition that estimates the probability of an event $A$ by retrieving
(some analogue of) a count of instances of $A$ from memory.  Such a
mechanism is necessarily subject to the requirements of set theory and
therefore implicitly embodies the equations of probability theory.

We expect this probability module to be based on observed event
frequencies, and to be unconscious, automatic, rapid, parallel,
relatively undemanding of cognitive capacity and evolutionarily `old'.
Support for this view comes from that fact that people make
probability judgments rapidly and easily and typically do not have
access to the reasons behind their estimations, from extensive
evidence that event frequencies are stored in memory by an automatic
and unconscious encoding process \citep{HasherZacks1984} and from
evidence suggesting that infants have surprisingly sophisticated
representations of probability \citep{cesana2012}.  Other support
comes from results showing that animals effectively judge
probabilities (for instance, the probability of obtaining food from a
given source) and that their judged probabilities are typically close
to optimal \citep{KheifetsGallistel2012}.

We also expect this probability module to be subject to occasional
intervention by other cognitive systems, and particularly by other
conscious and symbolic processes that may check the validity of
estimates produced.  We expect this type of intervention to be both
rare and effortful.  To quote one participant in an earlier experiment
where participants had to bet on a single event or on a conjunction
containing that event: `I know that the right answer is always to bet
on the single one, but sometimes I'm really drawn to the double one,
and it's hard to resist'.

\subsection{Comparing models of probabilistic reasoning}

The heuristics and biases approach proposes that people do not follow
the rules of probability theory when estimating probabilities: instead
they use various heuristics that sometimes give reasonable judgments
but sometimes lead to severe errors in estimation.  The results given
above directly contradict this proposal, showing that when bias due to
noise is cancelled, people's probability estimates closely follow the
fundamental rules of probability theory.  This cancellation of bias
cannot be explained in the heuristics view, because to know that the
bias in a given probabilistic expression should cancel requires access
to the rules of probability theory.  It is important to stress that
these results are the central point in our argument against the view
that people estimate probabilities via heuristics.  We are not arguing
that the heuristics and biases approach is incorrect because our
simple model of noisy retrieval from memory can explain four
well-known biases (there are many other biases in the literature which
our model does not address: base-rate neglect, the hard-easy effect,
confirmation bias, the confidence bias, and so on; see
\citet{Hilbert2012} for a review).  Instead, our main point is that
our experimental results demonstrate that the fundamental idea behind
the heuristics view (that people do not follow the rules of
probability theory) is seen to be incorrect when we use our simple
model to cancel the effects of noise.  Our results also argue against
models where people reason about probability using equations different
to those of probability theory
\citep{carlson1989disjunction,fantino1997conjunction,
  NilssonWinmanJuslin2009}.  Our results give support for models where
people follow probability theory in their probabilistic reasoning, but
are subject to the biasing effects of random noise; models such as
Minerva-DM \citep{MinervaDM} and Hilbert's `noisy channel' model
\citep{Hilbert2012}.
     
We presented our account under the assumption that, for any
event $A$, there is a clear-cut binary criterion for membership in
$A$: a given memory trace is either an instance of $A$ or it is
not.  Given the complexity of event representation and the graded
nature of most natural categories, this assumption is unrealistic: it
is more likely that stored instances vary in their degree of
membership in the category $A$, and that the process of retrieval from
memory would reflect this. 
Equally, our simple account assumes that
there is only one point at which random noise influences probability
estimation: the point at which memory is queried for stored instances
of event $A$.  Again, this is unrealistic: it is more likely that
noise has an influence on initial perceptual processes, on event
encoding, on retrieval, on subsequent processing and on
decision-making and action.  Further, our account applies only to unconditional
probabilities, not to conditional probabilities $P(B|A)$ (the
probability of $B$ given that $A$ has occurred).  An important aim for
future work is to see whether any useful predictions could be derived
by applying an extended version of our model to conditional
probabilities.

Clearly, a generalised version of our account, taking all of these
factors into account, would give give more a realistic description of
the processes of probability estimation.  This realism would come at
the cost of increased complexity, however: a more complex generalised
model would have various interacting components and parameters that
could be tuned in different ways to match behaviour.  Because of this
complexity, it would be difficult to derive clear and testable
predictions from such a model.  This is the main advantage of our
account: its simplicity allows us to derive clear, specific and
verifiable predictions about the impact of random variation on human
probabilistic reasoning.

\subsection{Concluding Remarks}

The focus in our work has been on people's estimation of simple,
unconditional probabilities.  Our results show that patterns of
systematic bias in these estimates can be explained via noise in
recall, and that when this noise is cancelled, people's estimates
match the requirements of probability theory closely, with no further
systematic bias.  This result has general implications for research on
people's use of heuristics in reasoning.  A frequent pattern in that
research is to identify a systematic bias in people's responses, and
to then take that bias as evidence that the correct reasoning process
is not being used.  We believe that this inference is premature: as we
have shown, random noise in reasoning can cause systematic biases in
people's responses even when people are using the correct reasoning
process.  To demonstrate conclusively that people are using
heuristics, researchers must show that observed biases cannot be
explained as the result of random noise.  To put it simply: biases do
not imply heuristics, and even a rational reasoning process can
produce systematically biased responses solely due to random variation
and noise.

\section*{Acknowledgements}

We would like to thank John Wixted, Mark Keane,  and 4 anonymous reviewers for their extremely helpful and insightful comments on earlier versions of this paper.

\clearpage

\begin{table}[t]
\label{Z1Z4}
	\caption{Average values of expressions $Z1_E, \ldots, Z4_E$ and
          difference from overall mean.}
	\begin{center}
 \begin{tabular}{| c| c| c| c|}
 \hline
 expression & M & SD & Diff.\ from overall mean (in units of SD) \\
 \hline
 $Z1_E$ & 25.37 & 	31.50 & 0.007 \\
 $Z2_E$ & 22.51	& 28.25 & -0.093\\
 $Z3_E$ & 26.50	& 27.95 & 0.048\\
 $Z4_E$ & 26.23 &	29.12 & 0.037 \\
 \hline
 Overall mean & 25.15 & 29.26 & \\
 \hline
\end{tabular} 
\end{center}
\end{table}

\clearpage

\begin{table}[t]
\label{Z5Z6}
	\caption{Average values of expressions $Z5_E,Z6_E$ and
          difference from predicted value of twice the overall mean of
          $Z1_E \ldots S4_E$, or $2 \times 25.15 =50.3$.}
\begin{center}
 \begin{tabular}{ |c |c| c| c|}
 \hline
 expression & M & SD & Diff.\ from $50.3$ (in units of SD) \\
 \hline
 $Z5_E$ & 48.74	& 42.24 &	-0.037 \\
 $Z6_E$ & 52.72	& 49.98	& 0.048 \\
 \hline
 overall mean & 50.73 &	46.30 & 0.001\\
 \hline
\end{tabular} 

\end{center}
\end{table}

\clearpage

\clearpage

\clearpage

\clearpage

\clearpage

\begin{figure}[t]
  \scalebox{.9}[.9]{
      \includegraphics*[viewport= 100 450 550 750]{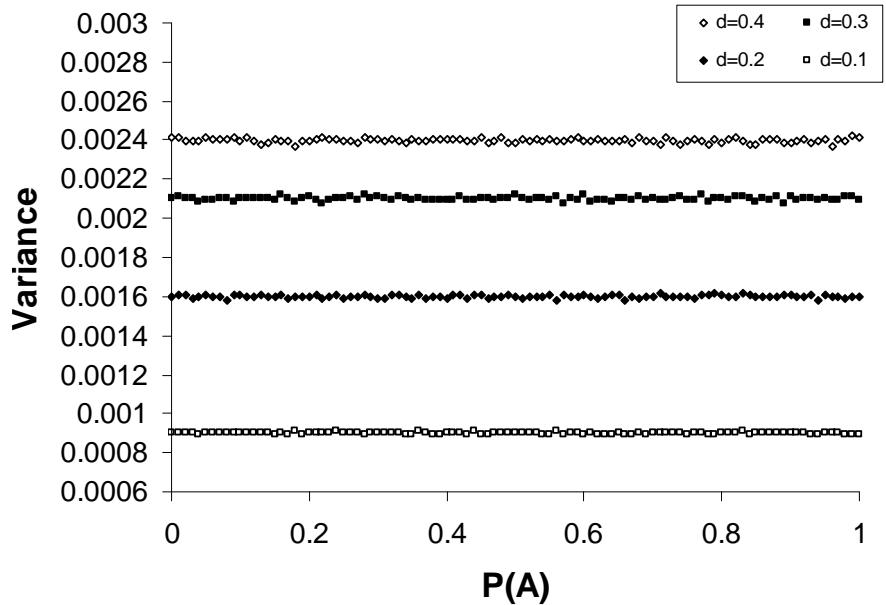}
     } 
\caption{
\label{simulatedVarianceGraph}
Computed variance in estimates $P_E(A)$ in Monte Carlo simulations with $m=100$,
for values of $P(A)$ from $0$ to $1$ in steps of $0.01$, and for $d=0.4$,
$0.3$, $0.2$ and $0.1$.  For each value of $P(A)$ and each value of
$d$, the simulation produced $10,000$ estimates $P_E(A)$: each point
in the graph represents the variance of one such set of $10,000$
estimates.  The critical point here is that this variance is
independent of $P(A)$ and, for a given value of $d$, is equal to the
predicted value $d(1-d)/m$.  For $d=0.4$, $0.3$, $0.2$ and $0.1$,
predicted variances are $0.0024$, $0.0021$, $0.0016$ and $0.0009$
respectively: the computed variances in the graph agree almost exactly
with those predictions. }
\end{figure}

\clearpage

\section*{{\center Appendix: \\ Variance in probability estimates in
the model}}

In our model, for any event $A$ the count $C(A)$ of the number of
flags in memory that are read as $1$ is made up of two components: (i)
the number of flags whose value is $1$ and which are read correctly as
$1$ and (ii) the number of flags whose value is $0$ but which are read
incorrectly as $1$.  Since the probability of a flag being read
incorrectly is $d$, the first component is a binomial random variable
with distribution $T_A-B(T_A,d)$ (where $T_A$ is the number of flags
whose true value is $1$ and $B(T_A,d)$ represents the binomal
distribution of the number of those flags that are incorrectly read)
and the second component is a binomial random variable with
distribution $B(m-T_A,d)$ (since there are $m- T_A$ flags whose true
value is $0$ and $B(m-T_A,d)$ represents the binomial distribution of
the number of those flags that are incorrectly read as $1$).  We thus
have
\begin{eqnarray*}
C(A) &=& T_A-B(T_A,d)+B(m-T_A,d)
\end{eqnarray*}
Since the mean of a binomial distribution $B(n,p)$ is $pn$, this gives
\begin{eqnarray*}
\overline{P_E(A)}& =& \frac{T_A -d T_A + d(m-T_A)}{m}\\ &=&
(1-2d)P(A)+d
\end{eqnarray*}
as in Equation \ref{eq:PEA}.  Since the expected variance of a binomal
$B(n,p)$ is $p(1-p)n$, the expected variances of these two
distributions are $d(1-d)T_A$ and $d(1-d)(m-T_A)$ respectively.  Since
each read of a flag is an independent Bernoulli trial with probability
$d$, these two distributions are independent.  For two independent
distributions the variance of their difference is equal to the sum of
their individual variances and so the expected variance of $C(A)$ is
\begin{eqnarray*}
V(C(A))&=& d(1-d)(T_A + m-T_A)\\
& =& d(1-d)m
\end{eqnarray*}
Finally, since  variance is defined as the average
squared difference from the mean, and since $P_E(A)=C(A)/m$, we get
\begin{eqnarray*}
V(P_E(A))& =& \frac{V(C(A))}{m^2} = d(1-d)/m
\end{eqnarray*}
and the expected variance in the distribution of $P_E(A)$ is a
constant $d(1-d)/m$ for all values of $P(A)$.

It is useful to check this result via simulation.  We did this using a
program that simulates the effects of random noise in recall on
probability estimations for a given set of probabilities (as in
Sections \ref{Expt1} and \ref{averagedEstimates}).  This program took
as input an event probability $P(A)$ and constructed a `memory'
containing $m=100$ items, each item containing a flag indicating whether
that item was an example of the given event.  The occurrence of those
flags in memory matched the given probability $P(A)$.  This
program also contained a noise parameter $d$; when reading flag values
from memory to generate an estimate $P_E(A)$, the program was designed
to have a random chance $d$ of returning the incorrect value.  For
each input probability $P(A)$ ranging from $0$ to $1$ in steps of
$0.01$ the program generated $10,000$ noisy estimates $P_E(A)$ and
used these $10,000$ values to estimate $V(P_E(A))$.

We carried out this simulation process for a range of values of $d$
($0.1$, $0.2$, $0.3$ $0.4$).  Figure \ref{simulatedVarianceGraph} graphs the average variance for these $10,000$ noisy estimates $P_E(A)$ for
each value of $P(A)$ and for each of those values of $d$.  As is clear
from the graph, the calculated  variance in the simulation was
independent of $P(A)$ and equal to $d(1-d)/m$, as expected.

\clearpage

\bibliography{SurprisinglyRationalReferences} 
\end{document}